\documentclass[10pt,preprint]{aastex}
\usepackage{listings}
\begin{document}

\title{Directionally Unsplit Hydrodynamic Schemes with Hybrid MPI/OpenMP/GPU Parallelization in AMR}
\shorttitle{Hybrid GPU Computing of Fluid in AMR}

\author{Hsi-Yu Schive\altaffilmark{1,2,3}\email{b88202011@ntu.edu.tw}, Ui-Han Zhang\altaffilmark{1,2,3}, \& Tzihong Chiueh\altaffilmark{1,2,3}}
\altaffiltext{1}{Department of Physics, National Taiwan University, 10617, Taipei, Taiwan}
\altaffiltext{2}{Center for Theoretical Sciences, National Taiwan University, 10617, Taipei, Taiwan}
\altaffiltext{3}{Leung Center for Cosmology and Particle Astrophysics (LeCosPA), National Taiwan University, 10617, Taipei, Taiwan}

\begin{abstract}
We present the implementation and performance of a class of directionally unsplit Riemann-solver-based hydrodynamic schemes on Graphic Processing Units (GPU). These schemes, including the MUSCL-Hancock method, a variant of the MUSCL-Hancock method, and the corner-transport-upwind method, are embedded into the adaptive-mesh-refinement (AMR) code GAMER. Furthermore, a hybrid MPI/OpenMP model is investigated, which enables the full exploitation of the computing power in a heterogeneous CPU/GPU cluster and significantly improves the overall performance. Performance benchmarks are conducted on the Dirac GPU cluster at NERSC/LBNL using up to 32 Tesla C2050 GPUs. A single GPU achieves speed-ups of 101(25) and 84(22) for uniform-mesh and AMR simulations, respectively, as compared with the performance using one(four) CPU core(s), and the excellent performance persists in multi-GPU tests. In addition, we make a direct comparison between GAMER and the widely-adopted CPU code Athena \citep{Stone2008} in adiabatic hydrodynamic tests and demonstrate that, with the same accuracy, GAMER is able to achieve two orders of magnitude performance speed-up.
\end{abstract}

\keywords{adaptive-mesh-refinement---graphic-processing-unit---hybrid MPI/OpenMP---hydrodynamics---methods: numerical}

\section{INTRODUCTION}

Manycore and accelerator-based programming models have become promising techniques in high-performance computing. Especially, novel uses of graphic processing units (GPU) with the parallel computing architecture CUDA \citep{NVIDIA2010} have revealed the potential of general-purpose GPU (GPGPU) computing. Moreover, nowadays the developments of GPU applications have moved beyond the single GPU stage, and both performance and parallel efficiency of the applications must be optimized. In astrophysical simulations, considerable performance speed-ups in multi-GPU systems have been demonstrated in a broad range of applications, for example, the direct $N$-body simulations \citep[e.g.,][]{Schive2008,Gaburov2009,Spurzem2011}, Barnes-Hut tree algorithm \citep{Hamada2009}, and reionization simulations \citep{AT2010}.

\citet{Schive2010a} present a parallel GPU-accelerated adaptive-mesh-refinement (AMR) code named \emph{GAMER}  (GPU-accelerated Adaptive-MEsh-Refinement), which is dedicated to high-performance and high-resolution astrophysical simulations. The AMR implementation is based on constructing a hierarchy of grid patches with an oct-tree data structure, and the relaxing total variation diminishing scheme \citep[RTVD;][]{JX1995} with directional splitting is adopted in the hydrodynamic solver. A hybrid CPU/GPU model is adopted in the code, in which both hydrodynamic and gravity solvers are implemented into GPU and the AMR data structure is manipulated by CPU. An order of magnitude performance speed-up is demonstrated on the \emph{Laohu} GPU cluster at the High Performance Computing Center at National Astronomical Observatories of China (NAOC), using up to 128 Tesla C1060 GPUs \citep{Spurzem2011}.

Directionally unsplit algorithms have the advantage of maintaining spatial symmetry and have been extended to magnetohydrodynamic (MHD) simulations, where the divergence-free constraint is preserved \citep[e.g.,][]{Stone2008}. However, due to the requirement of 3D stencils, it was unclear whether these schemes could be implemented into GPU with high performance. Accordingly, in this work, we describe the extensions to the original GAMER code, including the implementation of several directionally unsplit Riemann-solver-based hydrodynamic schemes, and the hybrid MPI/OpenMP/GPU parallelization. Benchmarks using up to 32 Tesla C2050 GPUs on the \emph{Dirac} GPU cluster at the National Energy Research Scientific Computing Center at Lawrence Berkeley National Laboratory (NERSC/LBNL) are reported.

The structure of this paper is as follows. In Section 2, we introduce the numerical algorithms of the directionally unsplit hydrodynamic schemes implemented in the code. In Section 3, we describe the GPU implementation and the performance comparison between CPU and GPU hydrodynamic solvers. Several optimizations, especially the hybrid MPI/OpenMP/GPU parallelization, are described in Section 4. In Section 5, we present the overall performance speed-ups in both uniform-mesh simulations and AMR simulations, using up to 32 Fermi GPUs. Finally, we summarize the results in Section 6.

\section{NUMERICAL ALGORITHMS}

In this work, three directionally unsplit hydrodynamic algorithms are implemented into GAMER, including the MUSCL-Hancock method \citep[MHM; see][for an introduction]{Toro2009}, a variant of the MUSCL-Hancock method \citep[hereafter referred to as the VL scheme;][]{Falle1991}, and the corner-transport-upwind scheme \citep[CTU;][]{Colella1990}. The three-dimensional CTU scheme adopted in the code is a simplified version proposed by \citet{GS2008}, which requires only six solutions to the Riemann problem per cell per time-step. Also note that the VL and CTU schemes have also been implemented in the widely-adopted code \emph{Athena} \citep{Stone2008,SG2009}, and hence enables a direct comparison between GAMER and Athena in terms of both accuracy and performance.

Since the CTU scheme is more complicated than MHM and VL, in the following we highlight the main procedures in CTU for updating solutions by one time step $\triangle t$, and emphasize the major differences between CTU and the other two schemes. For a more comprehensive description of the CTU implementation, please see \citet{Stone2008}.
\renewcommand{\labelenumi}{(\arabic{enumi})}
\begin{enumerate}
\item Evaluate the left and right interface values for all cell interfaces in all three spatial directions by the 1D spatial data reconstruction. For CTU, an intermediate step advanced by the characteristics is also included to evaluate the interface values at $\triangle t/2$.
\item Evaluate the fluxes across all cell interfaces by solving the Riemann problem.
\item Correct the half-step cell interface values obtained in step (1) by computing the transverse flux gradients, and solve the Riemann problem with the corrected data to obtain the new fluxes across all cell interfaces. This step is only necessary in CTU.
\item Update solutions by $\triangle t$ by the conservative integration.
\item Store the fluxes across the boundaries of all grid patches. This step is only necessary in AMR simulations.
\end{enumerate}

The characteristic tracing step and the evaluation of the transverse flux gradients are not required in either MHM or VL. In MHM, the half-step solutions are obtained by first computing the fluxes from the reconstructed interface values in step (1) without Riemann solvers and then advancing the interface values by $\triangle t/2$ by the flux differences in all three directions. In VL, the half-step solutions are obtained by computing the first-order fluxes with Riemann solver prior to step (1).

For the data reconstruction in step (1), GAMER supports both piecewise linear method (PLM) and piecewise parabolic method (PPM). In general, the latter is less diffusive and can resolve the density cusps in self-gravity systems with higher spatial resolution. Several slope limiters are implemented in both methods, including the generalized minmod limiter, van Leer-type limiter, and van Albada-type limiter \citep[see][for an introduction]{Toro2009}. We also support the hybrid limiter adopted in Athena, which combines the generalized minmod and van Leer-type limiters. The spatial interpolation can be applied to either primitive variables or characteristic variables. Performing interpolation on characteristic variables can reduce the non-physical oscillations in some cases, for example, in the shock tube problems. However, in cosmological simulations where low-density voids will always form, this method is found to be less robust and can yield negative density more easily.

The code supports several Riemann solvers, including the exact solver based on \citet{Toro2009}, HLLE solver \citep{Einfeldt1991}, HLLC solver \citep[see][for an introduction]{Toro2009}, and Roe's solver \citep{Roe1981}. The exact solver is much more time-consuming compared with other solvers and is implemented primarily for obtaining reference solutions. The HLLC and Roe's solvers give comparable accuracy, while the HLLE solver is more diffusive in circumstances requiring the information of contact waves. However, \citet{Einfeldt1991} showed that the HLLE solver is positively conservative; that is, the density and pressure are positive-definite. He also showed that in certain initial data the linearization approximation adopted in Roe's solver will fail and lead to negative density or pressure in the intermediate region of the Riemann-problem solution. If this failure is detected during simulations, we follow the same strategy adopted in Athena, in which either HLLE, HLLC, or exact solver is used to calculate the fluxes at the cell interfaces where the Roe's solver fails.

\section{GPU COMPUTING}

In this section, we describe the strategy to map the directionally unsplit hydrodynamic schemes described in Section 2 into the CUDA parallel computing architecture. We also present  the performance comparison between CPU and GPU solvers.

\subsection{CUDA Implementation}

In GAMER, the fluid data are always decomposed into grid patches, each of which consists of $8^3$ cells. In addition, due to the oct-tree data structure,  we can always group the eight nearby patches into a single \emph{patch group} (which contains $16^3$ cells) before updating data, so that the additional workload to prepare the ghost-cell data can be reduced in comparison with a single patch. In CUDA implementation, each patch group will be computed by one thread block. We do not require to store all simulation data in the GPU global memory. Instead, we only need to ensure that the workload in each GPU is high enough to fully exploit its computing power. Accordingly, the number of patch groups sent into GPU at a time depends on hardware specifications rather than simulation scale, and the maximum simulation scale will not be limited by the relatively small memory in GPUs (4 GB in Tesla C1060 and 3 GB in Tesla C2050).  Furthermore, since all simulation data still reside in the CPU memory, it reveals the potential of combining GPU computing with the out-of-core technique, by which the large storage space of multiple hard disks can be utilized as the additional virtual memory to significantly increase the total amount of memory available \citep{Schive2010b}.

The number of threads per thread block is a free parameter and the optimal value is also hardware-dependent. Typically, we use 128 threads per thread block in Tesla C1060 and 512 threads per thread block in Tesla C2050, in which cases each thread will compute the solutions of multiple cells. For a typical 3D loop in C language, we have
\begin{lstlisting}
// array format: Array3D[NZ][NY][NX]
for (unsigned int k=k_start; k<k_start+k_size; k++) {
for (unsigned int j=j_start; j<j_start+j_size; j++) {
for (unsigned int i=i_start; i<i_start+i_size; i++) {
   Array3D[k][j][i] = ... ;
}}}
\end{lstlisting}
, which is converted to the following GPU kernel:
\begin{lstlisting}
// array format: Array1D[NZ*NY*NX]
unsigned int count = threadIdx.x;
unsigned int blocksize = blockDim.x;
unsigned int i, j, k, index;
while (count<i_size*j_size*k_size) {
   i = i_start + count%i_size;
   j = j_start + count%(i_size*j_size)/i_size;
   k = k_start + count/(i_size*j_size);
   index = (k*NY + j)*NX + i;
   Array1D[index] = ... ;
   count += blocksize;
}
\end{lstlisting}
Note that additional workload, mainly associated with the expensive integer division and modulo operations, is introduced into the GPU kernel. However, it will not have a large performance impact as long as the number of arithmetic operations performed on the 1D array is large enough. Furthermore, since this kind of conversion poses no constraints on the number of threads per thread block, it makes the code much more flexible and also makes the performance tuning in different generations of GPUs more easily.

In GPU computing, one of the most important keys to achieving high performance is to efficiently utilize the small shared memory (16 KB per multiprocessor in Tesla C1060 and 48 KB per multiprocessor in Tesla C2050) in order to hide the latency of accessing data from the global memory. For the directionally split schemes, taking advantage of the shared memory is more straightforward and efficient since the integration only requires 1D stencils. Therefore, the data can be decomposed into many small 1D segments to fit into the shared memory. Moreover, the number of ghost cells in both ends of the 1D data segment can be much smaller than the total number of cells in one segment, and hence the computational overhead resulted from this 1D data decomposition is small. The shared memory has been fully utilized in the directionally split hydrodynamic schemes and the GPU Poisson solver in GAMER \citep{Schive2010a}.

In comparison, for the directionally unsplit hydrodynamic schemes described in Section 2, some calculations are essentially 3D operations which require 3D stencils (e.g., the steps (3) and (4) in CTU), while some are still 1D operations (e.g., the step (1) in CTU). For MHM, the half-step prediction also requires a 3D stencil since the solutions are updated by taking account of the flux differences in all three directions in a single step. Utilizing the shared memory in these 3D operations is more tricky, and the most straightforward solution is to decompose the data into small 3D tiles to fit into the shared memory. However, it is arguable whether this method can significantly improve the overall performance, especially when we consider the relatively high surface/volume ratio in each small 3D tile and the high arithmetic intensity in the Riemann-solver-based schemes. In some preliminary tests we find that no substantial performance improvement is obtained by using the shared memory. Therefore, in the current implementation, we only use the global memory for the directionally unsplit hydrodynamic schemes. Despite that, considerable performance speed-up  is still achieved and will be detailed in the next subsection. This approach makes the code more flexible to be run in GPUs with smaller shared memory (e.g., Tesla C1060), and it also makes it relatively straightforward to convert CPU hydrodynamic solvers to GPU solvers. Implementation of the shared memory in the directionally unsplit schemes will be investigated in the future. Also note that temporary global memory arrays are allocated to store the intermediate results during the integration (e.g., the fluxes and the left and right interface values). These arrays do not need to be transferred between host (CPU) memory and device (GPU) memory.

The integration procedure is carefully organized in order to minimize the performance impact of the high latency and relatively low bandwidth of the global memory access. For example, in the half-step prediction in MHM, the six interface values of one cell are all evaluated \emph{before} the thread proceeds to the next cell, so that the fluxes can be stored in the registers and the interface values can be updated without reloading data from the global memory. Another example is that the fluxes across the patch boundaries are stored immediately after solving the Riemann problem at these boundaries. By doing so, we do not need to perform additional memory copies in the global memory after all fluxes are evaluated, despite the fact that the latter method is more straightforward to implement.

In Fermi GPUs, the following tips are found to be able to fine-tune the performance. The compilation flag \emph{-Xptxas -dlcm=ca} is applied to enable both L1 and L2 caches. The flags \emph{-prec-div=false -ftz=true} are also used since they are found to improve the performance by 17\% without sacrificing the overall accuracy in most cases. During the initialization of CUDA devices, the cache configurations of the kernels of the directionally unsplit hydrodynamic solvers are set to \emph{cudaFuncCachePreferL1} via the function \emph{cudaFuncSetCacheConfig} in order to have a larger L1 cache. In a few cases, we find that the \emph{\underline{\ \ }forceinline\underline{\ \ }} function qualifier can enhance the performance, especially for computationally expensive functions (e.g., the Roe's Riemann solver). Additionally, it is observed that using structures instead of arrays to store the temporary 5-element fluid data delivers higher performance and also reduces the local memory usage in pre-Fermi GPUs.

Finally, we notice that many key components in different integration schemes are identical, for example, the data reconstruction, Riemann solver, full-step update, and storing fluxes. Therefore, in GAMER these functions are developed with flexible arguments so that they can be efficiently shared among different schemes. It makes the code more maintainable and extensible.

\subsection{GPU Performance}

All the tests in this work are performed on Tesla C2050 GPUs and Intel Xeon E5530 CPUs, except for the tests aiming to compare the performances between different hardware configurations. The CPU codes are compiled with \emph{gcc v4.4.2} and the GPU codes are compiled with \emph{nvcc v3.2}. The \emph{-O3} optimization flag is adopted in both CPU and GPU tests. Single precision variables are adopted in all performance experiments. Additionally, in order to have a fair comparison, the timing measurements of GPU always include the time to copy data between host and device memory.

\begin{figure}
\centering
\includegraphics[width=16cm]{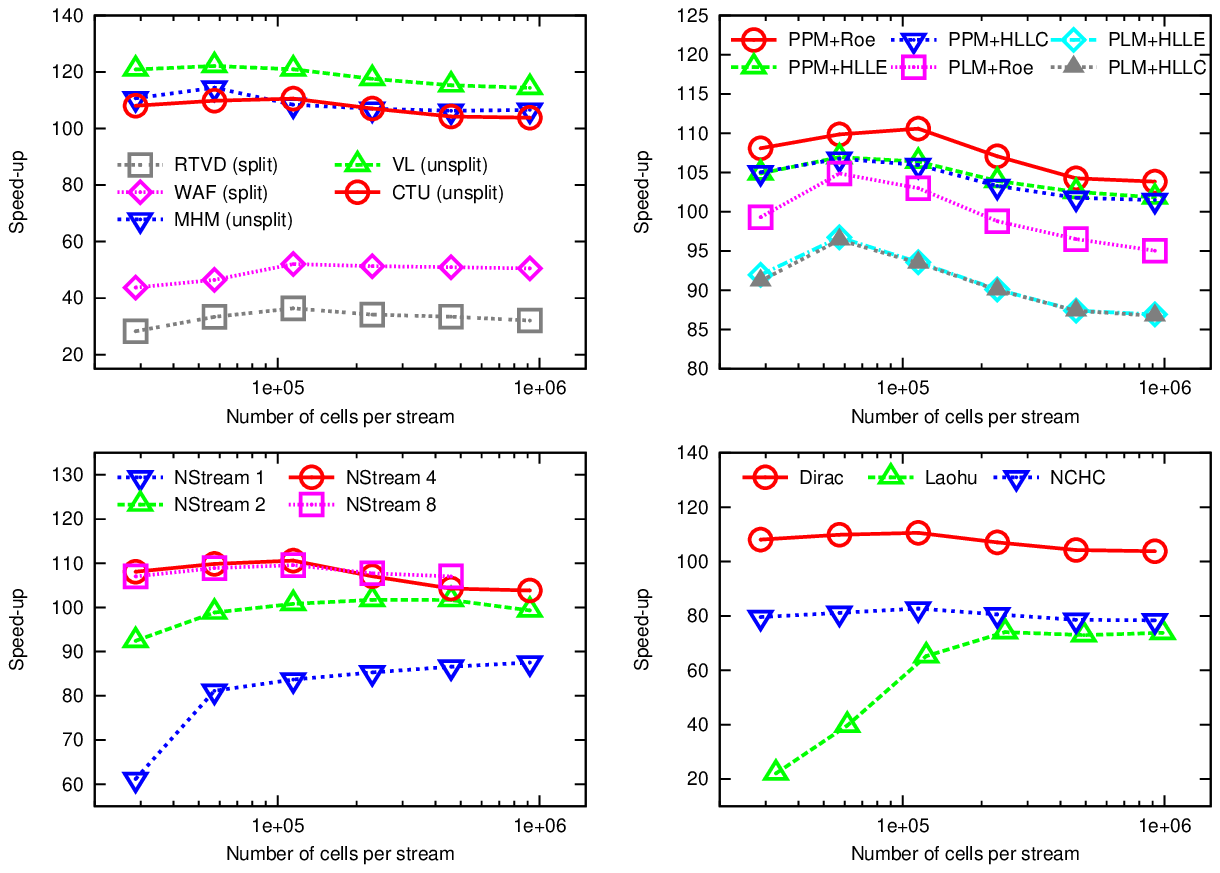}
\caption{Performance of the GPU hydrodynamic solvers. The top left panel shows the performances of different hydrodynamic schemes, including both the directionally unsplit schemes (MHM, VL, and CTU) with the PPM data reconstruction and Roe's Riemann solver and the split schemes (RTVD and WAF). The top right panel shows the performances of the CTU scheme with different data reconstruction methods (PLM and PPM) and different Riemann solvers (HLLE, HLLC, and Roe's solvers). The bottom left panel shows the performances of the CTU scheme with different numbers of CUDA streams. The bottom right panel shows the performances of the CTU scheme in different GPU systems (see text for details). The PPM data reconstruction and Roe's Riemann solver are adopted in the tests shown in the lower panels.}
\label{fig:Performance_GPUSolver}
\end{figure}

Figure \ref{fig:Performance_GPUSolver} shows the performance comparisons between one GPU and one CPU core as a function of the number of cells sent into GPU at a time. The top left panel shows the speed-ups of different hydrodynamic schemes. For comparison, we also include the results of two directionally split schemes: the RTVD scheme and the weighted average flux scheme \citep[WAF; see][for an introduction]{Toro2009}. Note that in the directionally split schemes we utilize the fast shared memory, while in the unsplit schemes we only use the global memory. Nevertheless, the latter still achieve substantially higher performance speed-ups due to their higher arithmetic intensity. Maximum speed-ups of 114, 122, and 111 are demonstrated in MHM, VL, and CTU, respectively. Also note that both VL and CTU require six Riemann solvers per cell per time-step, while MHM only requires three. Therefore, the MHM scheme is measured to be about 1.5 times faster than the other two schemes.

The \emph{CUDA stream} can be used to overlap the memory copy between host and device memory with the kernel execution, and, in general, using more CUDA streams can improve the efficiency of overlapping and hence deliver higher performance. The performances of the CTU scheme with different numbers of CUDA streams are given in the bottom left panel in Figure \ref{fig:Performance_GPUSolver}. The performance is improved by 26\% by using four CUDA streams as compared to the result using only one CUDA stream. It is also found that using more than four CUDA streams does not further improve the overall performance. Therefore, throughout the tests in this work, four CUDA streams are always adopted unless the number is explicitly specified. The method to implement the CUDA stream in GAMER is described in Section 4.1.

The top right panel in Figure \ref{fig:Performance_GPUSolver} compares the performances of the CTU scheme with different data reconstruction methods (PLM and PPM) and different Riemann solvers (HLLE, HLLC, and Roe's solvers). The highest speed-ups achieved in different methods range between 96 and 111, which do not differ significantly. In general, the PPM data reconstruction gives higher speed-ups than the PLM data reconstruction, and the Roe's solver delivers higher speed-ups than the HLLE and HLLC solvers. In addition, we find that the optimal performance is achieved when we have one to two patch groups per multiprocessor, in which case the CTU scheme with the PPM data reconstruction only requires about 110 MB to 220 MB memory in Tesla C2050. Having too many patch groups per multiprocessor can deteriorate the performance.

In order to have more comprehensive performance measurements, we also compare the timing results conducted in three different GPU systems: the \emph{Dirac} GPU cluster at NERSC/LBNL, the \emph{Laohu} GPU cluster at NAOC, and the GPU cluster at the National Center for High-performance
Computing of Taiwan (hereafter referred to as NCHC). The Dirac system contains 48 nodes connected by QDR InfiniBand. Each node is equipped with two Intel Xeon E5530 CPUs and one NVIDIA Tesla C2050 GPU. The Laohu system has 85 nodes, each with two Intel Xeon E5520 CPUs and two NVIDIA Tesla C1060 GPUs. The experimental GPU system in NCHC has two Intel Xeon X5670 CPUs and four NVIDIA Tesla M2070 GPUs in each node. The measured speed-ups in different systems are presented in the bottom right panel in Figure \ref{fig:Performance_GPUSolver}. The maximum number of cell updates per second achieved by one Fermi GPU is $3.0\times 10^{7}$, which is about 1.72 times faster than the performance achieved by on Tesla C1060 GPU.

Note that the timing experiments presented above only measure the performance of the hydrodynamic solver. In other words, they do not include the elapsed time of all other operations in AMR simulations. Therefore, the values in Figure \ref{fig:Performance_GPUSolver} can be considered as the optimal overall speed-ups we can achieve in any AMR simulation using GAMER. Also note that so far the performance comparisons are all based on one GPU versus one CPU core. The overall performance speed-up with multiple GPUs and CPU cores is presented in Section 5.

\section{OPTIMIZATION}

In this section, we describe several performance optimizations applied \emph{outside} the GPU kernel. Since GAMER adopts a hybrid CPU/GPU implementation, these optimizations have been demonstrated to be extremely important for achieving higher overall performance.

\subsection{Asynchronous Memory Copy}

As already mentioned in the previous section, the memory copy between host and device memory can be performed concurrently with the kernel execution by managing the CUDA stream. Specifically, the kernel launch and memory copy associated with different CUDA streams can be performed in parallel, while the operations associated with the same CUDA stream will be performed sequentially. The following code shows an example using $ns$ CUDA streams and assigning $np$ patch groups to each stream.
\begin{lstlisting}
//  input array format: float HostArray_In [ns*np][SizePerPatchGroup_In]
// output array format: float HostArray_Out[ns*np][SizePerPatchGroup_Out]
const unsigned int MemSize_In  = np*SizePerPatchGroup_In *sizeof(float);
const unsigned int MemSize_Out = np*SizePerPatchGroup_Out*sizeof(float);
for (unsigned int s=0; s<ns; s++) {
   cudaMemcpyAsync( DeviceArray_In + s*np, HostArray_In + s*np, MemSize_In,
                    cudaMemcpyHostToDevice, Stream[s] );
   Kernel <<< np, 512, 0, Stream[s] >>> ( DeviceArray_In + s*np, DeviceArray_Out + s*np, ... );
   cudaMemcpyAsync( HostArray_Out + s*np, DeviceArray_Out + s*np, MemSize_Out,
                    cudaMemcpyDeviceToHost, Stream[s] );
}
\end{lstlisting}

\subsection{Concurrent Execution between CPU and GPU}

In GAMER, before invoking the GPU solver, we need to prepare the input array in CPU, which stores the interior and ghost-cell data of the patch groups to be calculated. This step is referred as the \emph{preparation step}. In addition,  after receiving the updated data from GPU, we need to copy the data of different patches to their corresponding arrays in CPU. This step is referred as the \emph{closing step}. Experiments show that the preparation step can be very expensive, especially in AMR simulations where spatial and temporal interpolations are required to obtain the ghost-cell data if the targeted patches are adjacent to the coarse-fine boundaries. This step is the performance bottleneck in the previous version of GAMER and can take up to 3 times longer than the GPU computation.

To alleviate this issue, we notice that in principle the preparation and closing steps and the GPU solver can be performed concurrently, provided that they are targeting different patches. Therefore, by taking advantage of the fact that the GPU solver is an asynchronous function, we can overlap the executions of the preparation and closing steps in CPU with the GPU computation. The implementation and efficiency of this optimization are described in more detail in \citet{Schive2010a}.

\subsection{Hybrid MPI/OpenMP/GPU Parallelization}

The optimization of making CPU and GPU work in parallel greatly improves the overall performance, especially in the cases where the computation times of CPU and GPU are comparable (e.g., when relatively low-end GPUs are adopted). However, in both Dirac and Laohu systems, it does not completely eliminate the performance bottleneck described above due to the fact that the preparation step in CPU can take significantly longer than the GPU hydrodynamic solver. As a result, the overall performance speed-up is still much lower than the optimal value given in Section 3.2. To solve this issue, in this work we further implement the OpenMP parallelization in CPU computation.

In most GPU applications, one CPU core can take charge of one GPU, and the multi-GPU parallelization is achieved using MPI \citep[e.g.,][]{Schive2008,Schive2010a}. Accordingly, in the GPU clusters with more CPU cores than GPUs, some CPU cores will be idle during simulations. It is not an issue for the kind of GPU applications where the overall speed-up is dominated by the GPU performance. However, for complicated programs such as GAMER, the overall performance is determined by the performance of the hybrid CPU/GPU computing, and therefore it is crucial to fully exploit the computing power of both the multi-core CPUs and GPUs. To this end, we have implemented the hybrid MPI/OpenMP/GPU parallelization in GAMER. Each MPI process is responsible for one GPU, and the CPU computation allocated to each MPI process is further parallelized with OpenMP. For example, in a GPU cluster with $N_{node}$ nodes and each node is equipped with $N_{core}$ CPU cores and $N_{GPU}$ GPUs, we can run the simulation with $N_{node}\times N_{GPU}$ MPI processes and launch $N_{core}/N_{GPU}$ OpenMP threads in each process.

In GAMER, the AMR implementation is realized by constructing a hierarchy of grid patches, and different patches can be evaluated in parallel. Accordingly, the MPI implementation is based on the rectangular domain decomposition, and the OpenMP implementation is based on the patch-level parallelization. OpenMP is applied to most CPU computations, including both the preparation and closing steps, correcting the coarse-grid data, calculating the evolution time-step, and the grid refinement. Since different patches contain the same number of cells, the computation workload associated with each patch is approximately the same. Therefore, in most cases, OpenMP can be implemented straightforwardly and high parallel efficiency can be achieved.

\section{OVERALL PERFORMANCE}

In this section, we present the overall performance comparisons between CPUs and GPUs in both uniform-mesh simulations and AMR simulations. In each case, we compare the achieved speed-ups with different optimization levels described in Section 4, and show results for both single-GPU and multi-GPU. The CTU scheme with the PPM data reconstruction and Roe's Riemann solver is adopted in all tests presented in this section. Performance is measured on the Dirac system.

\subsection{Uniform-mesh Performance}

\begin{figure}
\centering
\includegraphics[width=12cm]{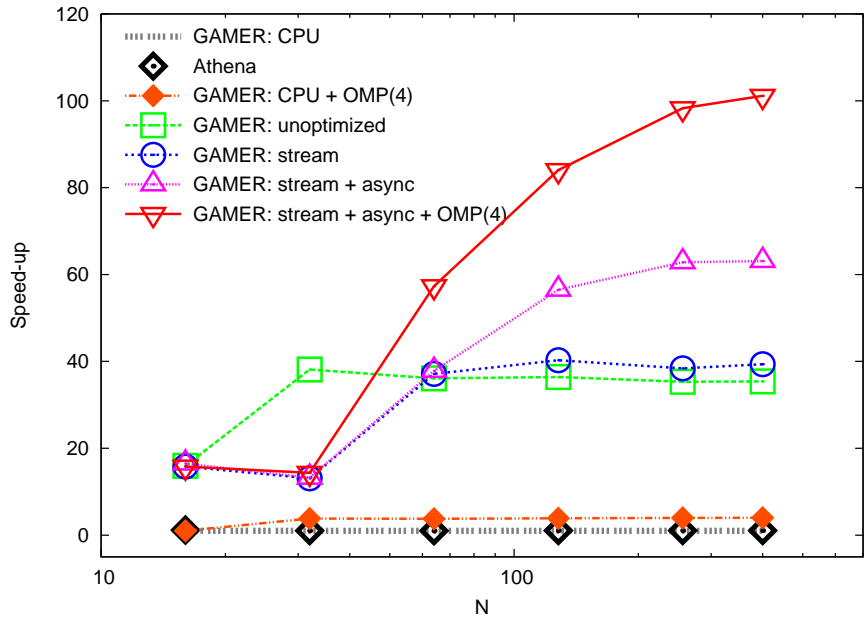}
\caption{Overall performance speed-up in single-GPU uniform-mesh simulations. The filled diamonds and thick dashed line show the CPU-only results with and without OpenMP, respectively. The unoptimized GPU performance is shown by the open squares. The open circles, triangles, and inverted triangles show the GPU performances, in which different optimizations are implemented successively. The abbreviation ``async'' represents the optimization of the concurrent execution between CPU and GPU. Four threads are used when OpenMP is enabled. We also show the results obtained by Athena (open diamonds) for comparison.  In all data points, the performance of GAMER using a single CPU core is regarded as the reference performance.}
\label{fig:Performance_GAMER_vs_Athena}
\end{figure}

Figure \ref{fig:Performance_GAMER_vs_Athena} shows the overall performance speed-up in the uniform-mesh 3D blast wave test as a function of the spatial resolution ($N$). First, we verify that the performance of GAMER without GPU acceleration and OpenMP is as fast as Athena. The performance difference is less than 6\% when $N\ge 32^3$. This result is very important as it makes the performance comparison between CPU and GPU in GAMER more convincing. The CPU-only performance scales linearly with the number of OpenMP threads when there are at least one patch groups per thread. For the performance with GPU acceleration, results with different optimization levels are shown together for comparison. The unoptimized code still achieves a speed-up of 35, and factors of 39, 63, and 101 speed-ups are further demonstrated when different optimizations described in Section 4 are implemented successively. This is a very encouraging result since the speed-up achieved by the fully optimized code closely approaches the optimal value given in Section 3.2. The maximum number of cell updates per second is $2.9\times 10^{7}$, which is 25 times faster than the CPU-only performance using four cores. We also compare the accuracies of physical results obtained separately by GAMER and Athena, and the relative differences are on the order of the machine precision in both single-precision and double-precision experiments.

Another important feature in GAMER is that it is very memory-efficient as compared with Athena; this is expected due to the patch-based decomposition employed in the code. For the directionally unsplit hydrodynamic schemes described in Section 2, usually we need to store the left and right interface values and fluxes at \emph{all} cell interfaces, which can be very memory-consuming. However, in GAMER we only need to store these data for the \emph{patches being computed in parallel}, the number of which is generally much smaller than the total number of patches. Experiments show that for the adiabatic hydrodynamic simulations with $N=400^3$, Athena consumes about 13 GB of memory, while GAMER only consumes roughly 3 GB of memory.

\begin{figure}
\centering
\includegraphics[width=12cm]{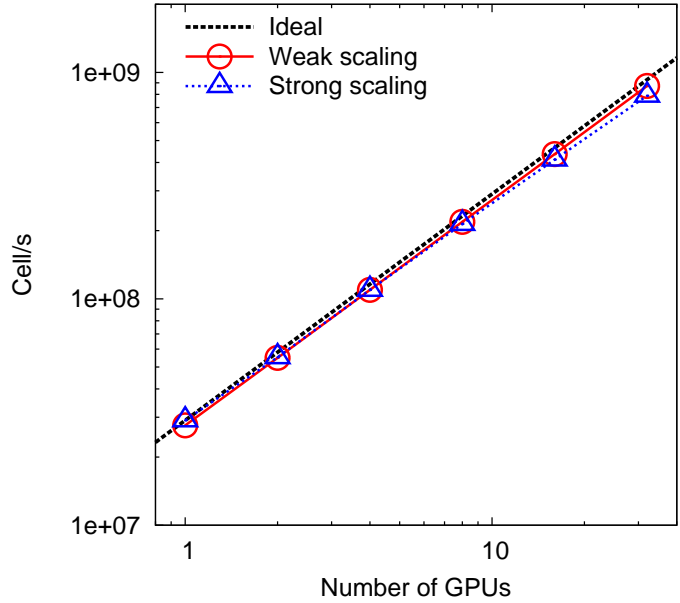}
\caption{Parallel scalability in uniform-mesh simulations. The open circles and triangles show the weak scaling and strong scaling, respectively. The ideal scaling is also shown for comparison (thick dashed line).}
\label{fig:Scalability_NoAMR}
\end{figure}

Figure \ref{fig:Scalability_NoAMR} shows the parallel scalability of GAMER in uniform-mesh simulations. The result is excellent for weak scaling, in which we let each GPU compute $512^3$ cells. The 32-GPU run is 31.6 times faster than the single-GPU performance, which corresponds to a parallel efficiency of 98.8\%. Strong scaling is much more challenging. By fixing the simulation resolution to $1024\times 1024\times 512$, the 32-GPU run still achieves a factor of 27.2 speed-up over the single-GPU performance, giving a parallel efficiency of 85.0\%.

\subsection{AMR Performance}

To demonstrate the performance in the AMR simulations with sufficiently complicated grid structure, we follow the similar timing experiments performed in \citet{Schive2010a} and measure the performance in purely baryonic cosmological simulations. The root-level resolution is $256^3$ and up to four refinement levels are used, giving $4096^3$ effective resolution. Note that the self-gravity is included here, and a GPU Poisson solver based on the successive overrelaxation method \citep[SOR; see][for an introduction]{Press2007} is adopted for the refinement levels. The GPU SOR solver alone is measured to be 84 times faster than the CPU counterpart on the Dirac system.

\begin{figure}
\centering
\includegraphics[width=12cm]{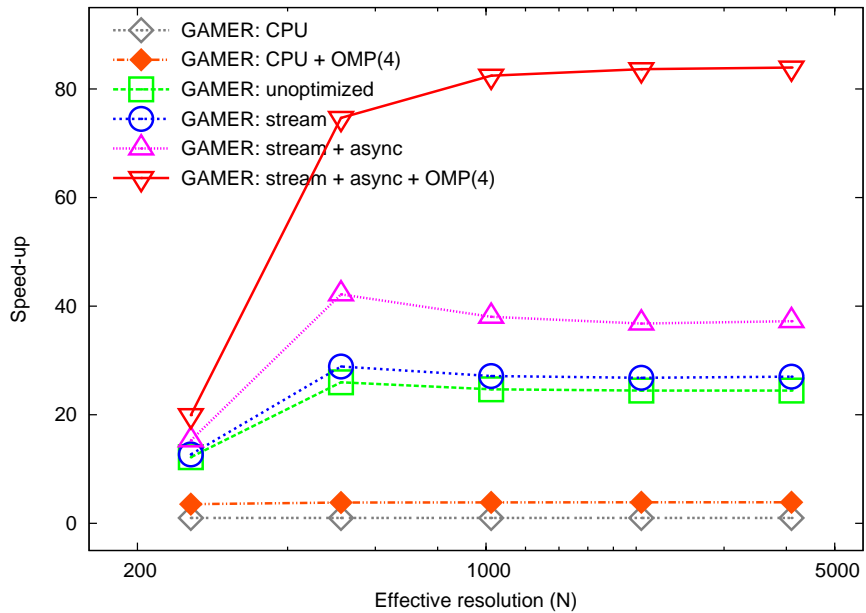}
\caption{Overall performance speed-up in single-GPU AMR simulations. Self-gravity is included in this test. The root-level resolution is $256^3$ and up to four refinement levels are used. Four threads are adopted when OpenMP is enabled. In all data points, the performance using a single CPU core is regarded as the reference performance.}
\label{fig:Performance_AMR_SingleGPU}
\end{figure}

Figure \ref{fig:Performance_AMR_SingleGPU} shows the overall performance speed-up at redshift $z=2$ using a single GPU. The fully optimized GPU code demonstrates speed-ups of 84.0 and 21.6 as compared with the CPU-only single-core and quad-core performances, respectively. The maximum speed-up is slightly lower than the value obtained in the uniform-mesh simulations, which is mainly due to the relatively lower speed-up ratio achieved by the GPU Poisson solver. We emphasize that the fully optimized performance (inverted triangles) is 2.3 times faster than the partially optimized performance without OpenMP (triangles), which demonstrates the importance of adopting the hybrid MPI/OpenMP parallelization. We also find that using more than four OpenMP threads does not further improve the overall performance, which is expected since the overall performance will be limited by the GPU performance when more than four OpenMP threads are used and the concurrent execution between CPU and GPU is enabled. This fact reveals the plausibility of installing more than one high-end GPUs in each multi-core computing node, such as the hardware configuration adopted on the Laohu system.

\begin{figure}
\centering
\includegraphics[width=12cm]{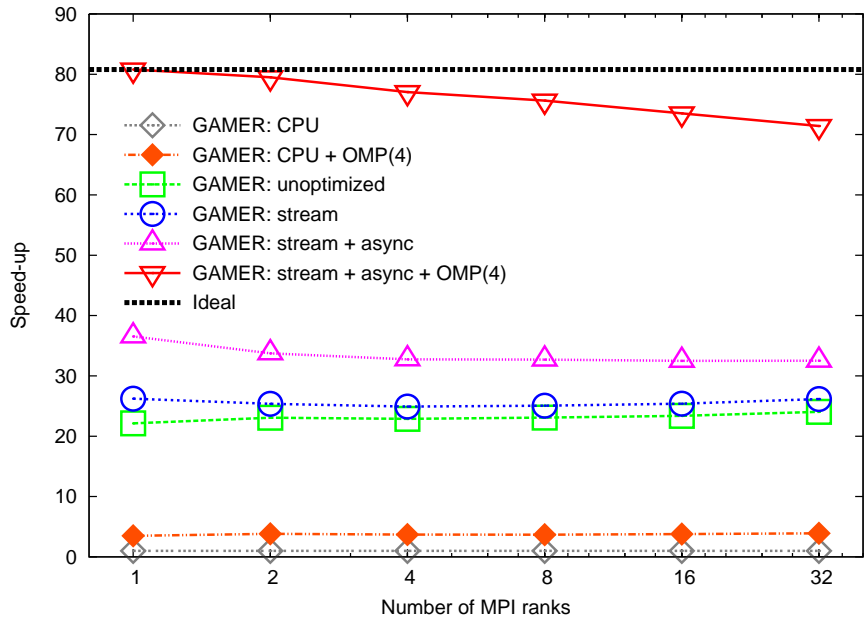}
\caption{Overall performance speed-up in multi-GPU AMR simulations as a function of the number of MPI ranks. The GPU performances with different optimization levels are compared to the CPU performance without OpenMP. For the CPU-only tests, the total number CPU cores is equal to the number of MPI ranks times the number of OpenMP threads per rank. For the GPU tests, the total number of GPUs is equal to the number of MPI ranks. Four threads per MPI rank are adopted when OpenMP is enabled.  The quad-core CPU performance (filled diamonds) and the ideal speed-up (thick dashed line) are also shown for comparison.}
\label{fig:Performance_AMR_MultiGPU}
\end{figure}

Figure \ref{fig:Performance_AMR_MultiGPU} shows the overall performance speed-up using multiple GPUs as a function of the number of MPI ranks. For example, the 8-GPU performance is compared to the CPU-only performance using 8 cores, each of which resides on a different computing node. In the 32-GPU test, the maximum speed-ups are 71.4 and 18.3 as compared with the CPU-only single-core and quad-core performances, respectively. The performance decrement results from the increasing MPI communication time, which takes about 11\% of the total execution time in the 32-GPU test. This issue was found to be of minor importance in the simulations with higher spatial resolution (and hence lower surface/volume ratio). For example, for the 128-GPU benchmark on the Laohu GPU cluster at NAOC, the MPI communication time takes less than 2\% of the total execution time \citep{Spurzem2011}. Also note that this issue can potentially be largely alleviated by overlapping communication with computation. Finally, we point out that, in the current implementation, the load is unbalanced among  different GPUs in the AMR simulations due to the rectangular domain decomposition. This issue will be addressed elsewhere.

\section{SUMMARY}

We have introduced the directionally unsplit hydrodynamic schemes newly implemented in GAMER, including the MUSCL-Hancock method, a variant of the MUSCL-Hancock method, and the corner-transport-upwind scheme. In each scheme, we support different data reconstruction methods (PLM and PPM), different Riemann solvers (HLLE, HLLC, and Roe's solvers), and also different slope limiters. All schemes have been implemented in GPU using NVIDIA CUDA, and up to two orders of magnitude performance speed-up has been demonstrated as compared to the performance using a single CPU core.

Several optimizations have been implemented in the code, including the asynchronous memory copy, the concurrent execution between CPU and GPU, and the hybrid MPI/OpenMP/GPU parallelization, by which we can fully exploit the computing power in a heterogeneous CPU/GPU system. OpenMP has been shown to be able to eliminate the performance bottleneck in the previous version of GAMER, and hence considerably improve the overall performance.

We have presented the overall performances of both uniform-mesh simulations and AMR simulations, measured on the Dirac cluster at NERSC/LBNL. In uniform-mesh tests, single GPU achieves a performance of $2.9\times 10^{7}$ cell updates per second, which is 101 times faster than the performance using a single CPU core and 25 times faster than the quad-core performance. We also directly compare GAMER with the well-known code Athena in adiabatic hydrodynamic tests, in which two orders of magnitude performance speed-up is also demonstrated. Weak scaling with 98.8\% parallel efficiency and strong scaling with 85.0\% parallel efficiency are achieved in the 32-GPU experiments. In AMR tests, 32-GPU run achieves speed-ups of 71.4 and 18.3 as compared to the performances using 32 and 128 CPU cores, respectively.

In Athena, both the VL and CTU schemes have been extended to MHD simulations \citep{GS2008,Stone2008,SG2009}. Following their work, we have performed some preliminary tests on a MHD solver with the CTU scheme and constrained transport technique, and achieved a factor of 60 speed-up when compared with a single CPU core. Details of the implementation of MHD in GAMER will be reported in a future communication.

\section{ACKNOWLEDGEMENTS}

A substantial part of simulations presented in this work were performed on the Dirac GPU cluster at the National Energy Research Scientific Computing Center at Lawrence Berkeley National Laboratory (NERSC/LBNL). We would like to thank Hemant Shukla, John Shalf, and Horst Simon in the International Center for Computational Science (ICCS) for providing the access to this system. The special supercomputer Laohu at the High Performance Computing Center at National Astronomical Observatories of China, funded by Ministry of Finance under the grant ZDYZ2008-2, has also been used. We want to thank Rainer Spurzem, Peter Berczik, and Gao Wei for helping conduct simulations on this system. Simulations were also performed on the National Center for High-Performance Computing of Taiwan. Finally, we are grateful to Evghenii Gaburov for insightful suggestions and sharing the source code. This work is supported in part by the National Science Council of Taiwan under the grant NSC97-2628-M-002-008-MY3.
.


\end{document}